\documentstyle[sprocl]{article}
\input{psfig}
\def\be{\begin{equation}}
\def\ee{\end{equation}}
\def\bea{\begin{eqnarray}}
\def\eea{\end{eqnarray}}

\def\bra{\langle}
\def\ket{\rangle}
\def\a{\alpha}
\def\b{\beta}
\def\g{\gamma}

\def\p{\pi}

\def\l{\lambda}
\def\m{\mu}
\def\n{\nu}
\def\G{\Gamma}

\def\to{\rightarrow}
\begin{document}
\thispagestyle{empty}
\begin{flushright}
DESY 97-144\\
SLAC-PUB-7612\\
ITP-SB-97-46\\
hep-ph/979708214\\
July 1997
\end{flushright}
\vspace*{1.0cm}
\title{
$B \to X_s \gamma$ 
in the Standard Model
\footnote{Invited talk presented by C.G. at the Second International
Conference on $B$ Physics and CP Violation, Honolulu, Hawaii,24-27 March
1997.}
}
\author{CHRISTOPH GREUB }
\address{
Deutsches Elektronen-Synchrotron, DESY, \\
Notkestr. 85, 22607 Hamburg,
Germany
}
\author{TOBIAS HURTH \footnote{Supported by Schweizerischer Nationalfonds
 and Department of Energy under contract number DE-ACO3-76SF00515}}
\address{Stanford Linear Accelerator Center,\\
 Stanford University, Stanford,
California 94309, USA\\
and\\
Institute for Theoretical Physics, SUNY at Stony Brook,\\
Stony Brook, New York 11794-3840, USA}

%
%
\maketitle\abstracts
{We give a summary about the various contributions which 
have to be calculated in order to obtain
the next-to-leading logarithmic
result for the branching ratio $BR(B \to X_s \gamma)$. 
Combining all these ingredients, which were obtained by different groups,
a complete next-to-leading-logarithmic  prediction of the inclusive 
decay rate was recently presented in the literature.
The theoretical uncertainty in the partonic decay rate is now 
at the 10\% level, i.e., less than
half of the error in the previous leading-logarithmic result.
We also mention the impact  of non-perturbative corrections which
scale like $1/m_b^2$ and discuss in some more detail the recently
discovered corrections which scale like $1/m_c^2$. 
It turns out that the $1/m_b^2$- and
the $1/m_c^2$ terms lead to corrections to the branching ratio
$BR(B \to X_s \gamma)$ well below the 10\% level.}
\section{Introduction}
The $B \to X_s \gamma$ decay has found increasing  attention over the last 
ten years. It provides an alternative approach in the search 
for physics beyond the standard model (SM). This decay, like other
rare $B$ meson decays, does not arise at the tree-level in the SM but is  induced 
by one-loop W-exchange diagrams, so nonstandard contributions 
(charged scalar exchanges, SUSY one-loop diagrams etc.) are not suppressed 
by an extra factor $\alpha/4\pi$ relative to the standard model amplitude. 
This high sensitivity for nonstandard contributions implies the possibility
for an indirect observation of new physics, a strategy complementary to the
direct production of new particles. The $B \to X_s \gamma$ decay plays 
already a very important role in  restricting the parameter space of 
extensions  of the SM like the minimal supersymmetric standard
model (MSSM) \cite{Bertolini,Wells}. 
However, even within the SM, the $B \to X_s \gamma$ decay  is  
important for constraining 
the Cabibbo-Kobayashi-Maskawa matrix elements involving the top-quark,
in particular $|V_{ts}|$. 
For both reasons, precise experimental and theoretical work
on this decay mode is required. 

On the theoretical side, the accuracy
in the dominating perturbative contribution 
was recently improved to next-to-leading precision 
\cite{AG91,Adel,Pott,GHW,Mikolaj,GGH}: The renormalization
scale dependence of the previous leading-log result at the $\pm 25\%$-level
was substantially reduced to $\pm 6\%$ and 
the central value was shifted out-side 
the $1\sigma$ bound of the CLEO measurement. Furthermore, 
the analysis of nonperturbtive 
contributions to the $B \to X_s \gamma$ decay mode was also recently improved:
The inclusive  $B \to X_s \gamma$ mode is 
theoretically much cleaner than the corresponding exclusive channels because
no specific model is needed
to describe the final hadronic state. According to Heavy Quark Effective 
Theory the class of 
non-perturbative effects which scales like $1/m_b^2$
is expected to be well below $10\%$ \cite{Falk}. This numerical
statement holds also 
for the recently discovered non-perturbative contributions
\cite{Voloshin,Wyler,Wise,Peccei,Rey}
which scale like $1/m_c^2$.
Thus the inclusive  $B \to X_s \gamma$ mode 
is well approximated by the partonic decay rate
$\G(b\to X_s \gamma)$ which can be
analyzed in renormalization group improved perturbation theory. 

Before reporting on these theoretical 
improvements in detail, we summarize the experimental status:
The observation of the exclusive $B \to K^* \gamma$
mode by CLEO \cite{Cleo1} in 1993 was  the first 
evidence for a penguin decay ever.  
An updated value \cite{Cleo3} for the branching 
ratio is  $BR(B \to K^* \gamma) =
(4.2 \pm 0.8 \pm 0.6) \times 10^{-5}$. 
In 1994 the CLEO collaboration
measured the inclusive  $B \to X_s \gamma$ branching ratio to be 
$(2.32 \pm 0.57 \pm 0.35) \times 10^{-4}$ 
where the first error is statistical and the second
is systematic \cite{Cleo2}: There are two separate CLEO analyses. 
The first one
measures the inclusive photon spectrum from B-decay near the end point.
The second technique constructs the inclusive rate by summing up the possible
exclusive final states. The branching ratio stated above is the average 
of the two measurements, taking into account the correlation between
the two techniques. 

There is also data from the LEP experiments:
While DELPHI \cite{DELPHI} in 1996  
and L3 \cite{L3} in 1993 have published the upper bounds
$BR(b \to s \gamma) < 5.4 \times 10^{-4}$ and 
$BR(b \to s \gamma) < 1.2 \times 10^{-3}$, respectively, the 
preliminary measurement
$BR(b \to s \gamma)=(3.38 \pm 0.74\pm 0.85)\times 10^{-4}$ 
by the ALEPH group was reported in the talk
by F. Parodi \cite{Parodi} at the 1997 Moriond meeting.
A similar number was also quoted by T. Skwarnicki \cite{Tomasz} 
in the heavy flavor meeting held in Santa Barbara in July 1997. 

More precise measurements are expected
from the upgraded CLEO detector, as well as from the B-factories
presently under construction at SLAC and KEK. 
In view of the expected high luminosity of the B-factories, experimental
accuracy of below $10\%$ appears to be in reach. 

The rest of the paper is organized as follows: Section 2
is devoted to 
the partonic (=perturbative) contribution to
$BR(B\to X_s \gamma)$. We explain in some detail the various 
calculational steps
leading to the next-to-leading logarithmic result.
In section 3 we briefly discuss the impact of 
the recently discovered non-perturbative
corrections which scale like $1/m_c^2$.

\section{Next-to-leading logarithmic corrections for $B \to X_s \gamma$}
It is well-known that the  QCD corrections enhance the 
partonic decay rate $ \Gamma(b \to s \g)$  by more than a factor of two. 
These QCD effects can be attributed to logarithms of the form 
$\alpha_s^n(m_b) \, \log^m(m_b/M)$,
where $M=m_t$ or $M=m_W$ and $m \le n$ (with $n=0,1,2,...$).
In order to get a reasonable result at all, one has  to sum at least
the leading-log (LL) series ($m=n$).  
Working to next-to-leading-log (NLL) precision means that 
one is also resumming all the
terms of the form $\a_s(m_b) \, \left(\a_s^n(m_b) \, \ln^n (m_b/M)\right)$.

An appropriate framework to achieve the necessary resummations 
is an  effective 
low-energy theory, obtained by integrating out the
heavy particles which in the SM are the top quark and the $W$-boson. 
The effective Hamiltonian relevant for $b \to s \gamma$ and
$b \to s g$ in the SM and many of its extensions reads 
\begin{equation}
\label{heff}
H_{eff}(b \to s \gamma)
       = - \frac{4 G_{F}}{\sqrt{2}} \, \lambda_{t} \, \sum_{i=1}^{8}
C_{i}(\mu) \, O_i(\mu) \quad ,
\end{equation}
where $O_i(\m)$ are the relevant operators,
$C_{i}(\mu)$ are the corresponding Wilson coefficients,
which contain the complete top- and W- mass dependence,
and $\lambda_t=V_{tb}V_{ts}^*$ with $V_{ij}$ being the
CKM matrix elements \footnote{The CKM dependence globally factorizes,
because we work in the approximation $\l_u=0$.}.
Neglecting operators with dimension $>6$ which are suppressed 
by higher powers of $1/m_{W/t}$ and using the equations
of motion for the operators, one arrives at the following basis 
of dimension 6 operators \cite{Grinstein90}
\bea
\label{operators}
O_1 &=& \left( \bar{c}_{L \b} \g^\m b_{L \a} \right) \,
        \left( \bar{s}_{L \a} \g_\m c_{L \b} \right)\,, \nonumber \\
O_2 &=& \left( \bar{c}_{L \a} \g^\m b_{L \a} \right) \,
        \left( \bar{s}_{L \b} \g_\m c_{L \b} \right) \,,\nonumber \\
O_7 &=& (e/16\p^{2}) \, \bar{s}_{\a} \, \sigma^{\m \n}
      \, (m_{b}(\mu)  R + m_{s}(\mu)  L) \, b_{\a} \ F_{\m \n} \,,
        \nonumber \\
O_8 &=& (g_s/16\p^{2}) \, \bar{s}_{\a} \, \sigma^{\m \n}
      \, (m_{b}(\mu)  R + m_{s}(\mu)  L) \, (\l^A_{\a \b}/2) \,b_{\b}
      \ G^A_{\m \n} \quad .
\eea
Because the Wilson coefficients of the penguin induced four-fermion
operators $O_3,...,O_6$ are very small, we do not list them here.
In this framework the next-to-leading logarithmic terms 
$\a_s(m_b) \, (\a_s^n(m_b) \log^n(m_b/m_{W/t}))$ 
in the $b \to s \gamma$ amplitude
have
two sources:\\ 
{\bf 1} $\bullet$ The NLL Wilson 
coefficients $C_i(\mu)$ at the 
scale $\mu \approx m_b$ contain leading and next-to-leading logarithmic
terms in resummed form. \\
{\bf 2} $\bullet$ The $O(\a_s)$ 
corrections to the matrix elements of the operators $O_i$ yield
next-to-leading order terms
when multiplied by the (leading logarithmic part of the) Wilson 
coefficients. 
\\
We stress that only the sum of these two sources
is {\it independent of the renormalization scheme}. 
Let us discuss in some more detail the contributions mentioned
in ${\bf 1}$ and ${\bf 2}$: \\ 

{\bf ad 1} $\bullet$  
{}From the $\mu$-independence of the effective Hamiltonian,
one can derive a renormalization group equation 
(RGE) for the Wilson 
coefficients $C_i(\mu)$:
\be
\label{RGE}
\mu \frac{d}{d\mu} C_i(\mu) = \gamma_{ji} \, C_j(\mu) \quad ,
\ee  
where the $(8 \times 8)$ matrix $\gamma$ is the anomalous dimension
matrix of the operators $O_i$.
To solve this first order differential equation one explicitly needs 
initial conditions $C_i(\mu_0)$ at some scale $\mu_0$ 
as well as the anomalous
dimension matrix $\gamma_{ij}$.\\
{\bf 1a:}  
The initial conditions are obtained
by matching the effective theory  
to the full standard model theory
at the scale $\m_0=\m_W$, where
$\m_W$ denotes a scale of order $m_W$ or $m_t$. At this scale,
the matrix elements of the operators  in the 
effective theory lead to the  same logarithms  as the full theory
calculation. 
Consequently, the Wilson coefficients 
$C_i(\m_W)$ only pick up small QCD corrections,
which can be calculated in fixed-order perturbation theory.
In the LL (NLL) program, the matching has to be worked out to order
$\a_s^0$ ($\a_s^1$) precision. 
\\
{\bf 1b:} 
Solving the RGE (\ref{RGE}) and using the $C_i(\m_W)$ 
of Step ${\bf 1a}$ as initial conditions, one performs the   
evolution of these Wilson coefficients from 
$\m=\m_W$ down to $\m = \m_b$, where $\m_b$ is of the order of $m_b$.
As the matrix elements of the operators evaluated at the low scale
$\m_b$ are free of large logarithms, the latter are contained in resummed
form in the Wilson coefficients. For a LL (NLL) calculation, this RGE step
has to be performed using the anomalous dimension matrix $\gamma_{ij}$  up 
to order $\a_s^1$ ($\a_s^2$).\\
{\bf ad 2} $\bullet$ The matrix elements 
of the operators $\bra s \g |O_i (\mu)|b \ket$ at the scale  $\mu = \m_b$
have to be calculated to order $\a_s^0$ ($\a_s^1$) in the LL (NLL)
calculation.

\begin{figure} 
\centerline{
\psfig{figure=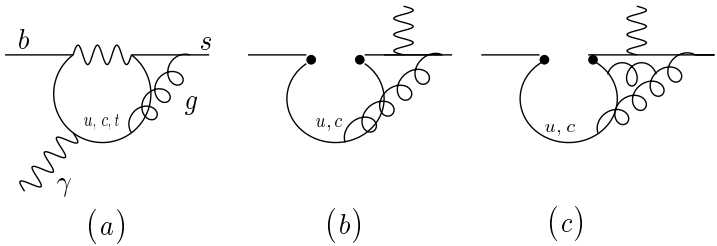,height=2.5in}} 
\caption[]{a) Typical diagram (full theory) contributing 
in the NLL matching calculation. b) Typical diagram  contributing to
the matrix element of the operator $O_2$.
c) Typical contribution to the $O(\a_s^2)$ anomalous dimension matrix.
\label{feynman}}
\end{figure}

Until recently, only the leading logarithmic (LL) perturbative QCD
were known systematically 
\cite{counterterm}. The error in this approximation 
was dominated by a large 
renormalization scale dependence at the $\pm 25\%$ level. 
The measurement of the CLEO collaboration \cite{Cleo2} overlaps with 
the estimates based on leading logarithmic calculations
(or with some next-to-leading effects partially included)
and the experimental and 
theoretical errors are comparable 
\cite{AG91,Ciuchini,Burasno,Shifman}. 
However, in view of the expected 
increase in the experimental precision in the near future, 
it became clear that 
a systematic inclusion of the NLL corrections
was necessary.  
This ambitious NLL enterprise was recently completed. 
All three steps ({\bf 1a,1b,2}) involve rather difficult
calculations. The most difficult part in Step {\bf 1a} is the 
two-loop (or  
order $\a_s$) matching of the dipole operators $O_7$ and $O_8$. 
It involves two-loop
diagrams both in the full and in the effective theory 
(see Fig. \ref{feynman}a). 
This matching calculation was done by Adel and Yao \cite{Adel} some time ago. 
As this is a crucial step in the NLL program,
Greub and Hurth \cite{GGH}
recently confirmed their findings in a detailed 
re-calculation, using a somewhat different method.
In order to match dimension 6 operators $O_7$ and $O_8$, 
it is sufficient to
extract the terms  of order $m_b \, \frac{m_b^2}{M^2} $ ($M=m_W,m_t$)
from the standard model 
matrix elements for $b \to s \g$ and $b \to s g$.
Terms supressed by additional powers of
$m_b/M$ correspond to higher
dimensional operators in the effective theory.
In \cite{GGH} the finite parts of the 
two-loop diagrams in the SM were calculated by means of the 
well-known Heavy Mass Expansion (HME) which naturally leads 
to a systematic expansion of Feynman diagrams in inverse powers of
$M$. We mention here that the evolution of the Wilson coefficients 
between $\mu=m_t$ and $\mu=m_w$ to LL precision
implied  an additional contribution of $+10\%$ to the leading-log prediction
for the decay rate \cite{Cho}. 
Most of this contribution is automatically
included in the NLL matching at the $m_W$-scale \cite{Adel,GGH}, because 
the first term of the LL-sum of \cite{Cho} is reproduced  
and higher order terms
$(\alpha_s \log(\frac{m_t}{m_W}))^n ( n>1)$ are rather small. 
In addition, the NLL matching result includes  
the first term of the NLL-sum. 
    
Step {\bf 2} basically consists of Bremsstrahlung corrections and virtual
corrections. While the Bremsstrahlung corrections
(together with some virtual corrections needed to cancel
infrared singularities) were worked
out some time ago by Ali and Greub \cite{AG91} and have  
been confirmed and extended by Pott \cite{Pott}, a  
complete analysis of the virtual corrections (up to the contributions 
of the four-fermion operators with very small coefficients) was presented
by Greub, Hurth and Wyler \cite{GHW}. This calculation also involves two-
loop diagrams where the full charm quark mass
dependence has to be taken into account. A typical diagram is shown
in Fig. \ref{feynman}b.
By using Mellin-Barnes techniques in the Feynman parameter integrals, 
the result of these two-loop
diagrams was obtained in the form
\be
\label{Mellin}
c_0 + \sum_{n=0,1,2,...;m=0,1,2,3} c_{nm} \left( \frac{m_c^2}{m_b^2}
\right)^n \, \log^m \frac{m_c^2}{m_b^2} \quad ,
\ee
where the quantities $c_0$ and $c_{nm}$ are independent of $m_c$.
Note, that a finite result is obtained in the limit $m_c \to 0$,
as there is no naked logarithm of $m_c^2/m_b^2$. This observation
is of some
importance in the $b \to d \gamma$ process, where the $u$-quark
propagation
in the loop is not CKM suppressed. It is, however, even more important 
that the inclusion of the $O(\a_s)$ matrix elements
leads to a drastic reduction of the renormalization 
scale uncertainty from about $\pm 25\%$ to about $\pm 6\%$.
Analytically, the reason is, that the term $\a_s \log(\mu/m_b)$ which
dominates the $\mu$-dependence of the LL result, is cancelled by
a corresponding term appearing in the $O(\a_s)$ matrix element.
Finally, the anomalous dimension matrix (at $O(\a_s^2)$), Step {\bf 1b}, 
has been worked out
by Chetyrkin, Misiak and M\"unz \cite{Mikolaj}. 
The calculation of the elements $\gamma_{i7}$ 
and $\gamma_{i8}$ ($i=1,...,6$) in the $O(\a_s^2)$ anomalous
dimension matrix involves a huge number of three loop-diagrams
from which the pole parts (in the $d-4$ expansion) have to be extracted.
For a typical diagram see Fig. \ref{feynman}c. The extraction of the pole
parts were simplified by a clever decomposition of the scalar propagator.
Moreover, the number of necessary evanescent operators were reduced by 
a new choice of a basis of dimension 6 operators.     
Using the matching result (Step {\bf 1a}), these authors obtained 
the next-to-leading correction to the Wilson coefficient $C_7(\m_b)$
which is the only relevant one for the $b \to X_s \gamma$ decay rate.  
Numerically, the LL and the NLL values 
for $C_{7}(\m_b)$ 
are rather similar; the NLL 
corrections to the Wilson coefficient $C_7(\m_b)$ 
lead to a change of the  $b \to X_s \gamma$
decay rate which does not exceed  $\pm 6\%$ \cite{Mikolaj}: 
 The new contributions can be split into a part
which is due to the order $\a_s$ corrections to the matching (Step 1a) 
and into
a part stemming from the improved anomalous dimension matrix (Step 1b). 
While individually these two parts are not so small (in the
NDR scheme, which was used in \cite{Mikolaj}), they almost cancel
when combined as illustrated in \cite{Mikolaj}. 
This shows that all the three different pieces, {\bf 1a,1b,2}, are numerically
equally important. 

Combining the NLL calculations of all the three steps ({\bf 1a+b,2}), 
the first complete theoretical prediction to NLL  pecision 
for the $b \to X_s + \gamma$ branching ratio 
was presented in \cite{Mikolaj}:
$BR(B \to X_s \g)=(3.28 \pm 0.33) \times 10^{-4}$.  
The error is due to the $\pm 6\%$ renormalization scale uncertainty
and due to the $\pm 8\%$ combined uncertainty in the input parameters.

\section{$1/m_b^2$ and $1/m_c^2$ corrections}
Neglecting perturbative QCD corrections and assuming that 
$B \to X_s \gamma$ is due to the operator $O_7$ only,
the calculation of the differential decay rate basically amounts
to work out the imaginary part of the forward scattering
amplitude $T(q)$
\be
\label{forward}
T(q) = i \, \int d^4x \, \bra B|T O_7^+(x) \, O_7(0) |B \ket \,
\exp (iqx) \quad .
\ee
Using the operator product expansion for $T O_7^+(x) \, O_7(0)$
and Heavy Quark Effective Theory methods, the decay width
$\Gamma(B \to X_s \gamma)$ reads \cite{Falk} (modulo higher terms in the
$1/m_b$ expansion)
\bea
\label{width}
\Gamma_{B \to X_s \gamma}^{(O_7,O_7)} &=&
\frac{\a G_F^2 m_b^5}{32 \pi^4} \, |V_{tb} V_{ts}|^2 \, C_7^2(m_b) \,
\left( 1 + \frac{\delta^{NP}_{rad}}{m_b^2} \right) \quad , \nonumber \\
\delta^{NP}_{rad} &=& \frac{1}{2} \l_1 - \frac{9}{2} \l_2 \quad ,
\eea
where $\l_1$ and $\l_2$ are the kinetic energy- and the chromomagnetic
energy parameters. Using $\l_1=-0.5 \, GeV^2$ and $\l_2=0.12 \, GeV^2$,
one gets $\delta_{rad}^{NP} \simeq -4\%$.
As also the semileptonic decay width gets $1/m_b^2$ corrections
which are negative (see e.g. \cite{Manohar}), 
these non-perturbative corrections tend to cancel
in the branching ratio $BR(B \to X_s \gamma)$ and only about $1\%$
remains. This contribution was already included in the 
theoretical NLL prediction presented in section 2 of this article.

\begin{figure} 
\centerline{
\psfig{figure=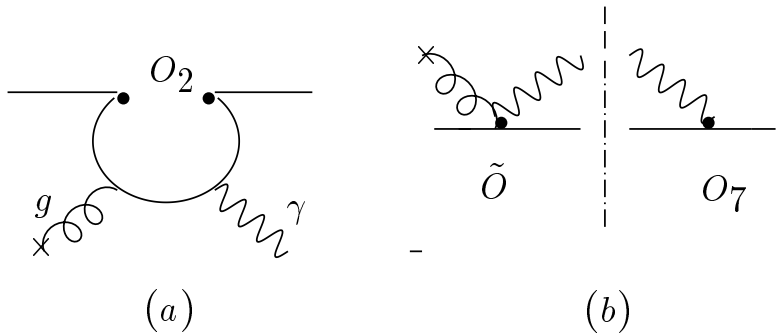,height=2.5in}} 
\caption[]{a)Feynman diagram from which the operator $\tilde{O}$
arises. b) Relevant cut-diagram for the $(O_2,O_7)$-interference.
\label{Voloshinfig}}
\end{figure}

Recently, Voloshin \cite{Voloshin} considered the non-perturbative
effects when including also the operator $O_2$. 
This effect is generated from the diagram in Fig. 
\ref{Voloshinfig}a 
(and from the one not shown where the gluon and the photon
are interchanged); $g$ is a soft gluon interacting with the
charm quarks in the loop. Up to a characteristic Lorentz structure,
this loop is given by the integral
\be
\label{volloop} 
\int_0^1 dx \, \int_0^{1-x} dy \, \frac{xy}{m_c^2-k_g^2 x(1-x) -2xy k_g k_\g}
\quad .
\ee
As the gluon is soft, i.e., $k_g^2,k_g k_\g \approx \Lambda^{QCD} \, m_b/2
\ll m_c^2$, the integral can be expanded in $k_g$. The (formally)
leading operator, denoted by $\tilde{O}$, is
\be
\label{tildeo}
\tilde{O} = \frac{G_F}{\sqrt{2}} V_{cb} V_{cs}^* C_2 \,
\frac{e Q_c}{48 \pi^2 m_c^2} \, \bar{s} \g_\mu (1-\g_5) g_s 
G_{\nu \lambda} b \, \epsilon^{\mu \nu \rho \sigma} \partial^\lambda
F_{\rho \sigma} \quad .
\ee 
Working out then the cut diagram shown in Fig. \ref{Voloshinfig}b,
one obtains the non-perturbative 
contribution $\Gamma^{(\tilde{O},O_7)}_{B \to X_s \gamma}$
to the decay width,
which is due to the $(O_2,O_7)$ interference.
Normalizing this contribution by the LL partonic width, one obtains
\be
\label{voleffect}
\frac{\Gamma^{(\tilde{O},O_7)}_{B \to X_s \gamma
}}{\Gamma_{b \to s \g}^{LL}} = -\frac{1}{9} \, \frac{C_2}{C_7} 
\frac{\l_2}{m_c^2} \simeq +0.03 \quad .
\ee
Including this correction with the sign found in \cite{Rey}, the NLL
prediction for the branching ratio becomes
$BR(B \to X_s \g)=(3.38 \pm 0.33) \times 10^{-4}$.  

As the expansion parameter is $m_b \Lambda_{QCD}/m_c^2 \approx 0.6$
(rather than $\Lambda^2_{QCD}/m_c^2$), it is not a priori clear
whether formally higher order terms in the $m_c$ expansion are
numerically suppressed. More detailed investigations 
\cite{Wise,Peccei,Rey}
show that higher order terms are indeed suppressed, because
the corresponding expansion coefficients are small.

We mention that the analogous $1/m_c^2$ effect 
has been found independently in the exclusive mode
$B \to K^* \gamma$ in ref. \cite{Wyler}. Numerically, the effect
there is also at the few percent level.

\section{Summary}
Collecting all NLL contributions and the small nonperturbative 
correction which scales with $m_b^2$,  the final analysis done 
by Chetyrkin, Misiak and M\"unz yields
$BR(B \to X_s \g)=(3.38 \pm 0.33)\times 10^{-4}$ 
when also the $+3\%$ shift due to the
non-perturbative effects from the $1/m_c^2$
corrections is included.  
The theoretical error in the NLL prediction is reduced by a factor of 2
when compared with the LL result. 
This theoretical value for the branching ratio is 
in agreement with the CLEO
measurement (at the $2\sigma$-level) and also with the recent (preliminary)
measurement by ALEPH.
Clearly, the inclusive $B \to X_s + \gamma$ 
mode will provide an interesting test of the SM and its extensions
as soon as more precise experimental data  become  available. \\

{\bf Note added:} 
When finishing this article, we received the new 
work by Buras, Kwiatkowski and Pott \cite{Kwiatkowski}.
While these authors fully confirm the matching conditions by
\cite{Adel,GGH}, their analysis is slighly different, leading
to the branching ratio 
$BR(B \to X_s \g)=(3.48 \pm 0.31)\times 10^{-4}$. The shift in the
central value is due to systematically discarding next-next-leading 
order terms, while in the earlier analysis \cite{GHW,Mikolaj} some terms 
of this order were
 included. Also their estimate for the remaining renomalization
scale dependence is somewhat different: The $\mu$-uncertainties in the
decay width for the radiative decay and the semileptonic decay were
treated independently and added in quadrature. In the old analysis
\cite{GHW,Mikolaj} the scale $\mu$ was varied simultaneously in both decays. 
As the semileptonic decay width is an increasing function of $\mu$ while
the radiative decay width is decreasing, a larger $\mu$-uncertainty was 
obtained which {\it as a more} conservative estimate  we finally prefer.
The results are fully compatible after all.\\

{\bf Acknowledgements:}
We thank G. Buchalla, A.Buras, R. Forty and T. Swarnicki
for discussions. C.G. thanks the organizers for the great week
on Oahu.

\end{document}